\documentclass[epj]{webofc}
\usepackage[utf8]{inputenc}
\usepackage[varg]{txfonts}   % Web of Conferences font
\usepackage{booktabs}
\usepackage{xcolor}
\definecolor{darkred}{rgb}{0.4,0.0,0.0}
\definecolor{darkgreen}{rgb}{0.0,0.4,0.0}
\definecolor{darkblue}{rgb}{0.0,0.0,0.4}
\usepackage[bookmarks,linktocpage,colorlinks,
    linkcolor = darkred,
    urlcolor  = darkblue,
    citecolor = darkgreen]{hyperref}
\usepackage{slashed}
\usepackage{subfigure}
\usepackage{amsmath,amssymb}
\usepackage{bbm}
\newcommand{\beq}{\begin{equation}}
\newcommand{\eeq}{\end{equation}}
\newcommand{\bea}{\begin{eqnarray}}
\newcommand{\eea}{\end{eqnarray}}
\newcommand{\eq}[1]{eq.(\ref{eq:#1})}
\newcommand{\re}{{\rm Re}}
\newcommand{\im}{{\rm Im}}
\wocname{EPJ Web of Conferences}
\woctitle{Lattice2017}
\begin{document}
%%%%%%%%%%%%%%%%%%%%%%%%%%%%%%%%%%%%%%%%%%%%%%%%%%%%%%%%%%%%%%%%%%%%%%%%%%%%%
%
\selectlanguage{english}
%----------------------------------------------------------------------------
\title{%
A complex path around the sign problem
}
%----------------------------------------------------------------------------
\author{%
\firstname{Paulo F. }  \lastname{Bedaque}\inst{1}\fnsep\thanks{Speaker, \email{bedaque@umd.edu} }
% etc.
}
%----------------------------------------------------------------------------
\institute{%
University of Maryland, College Park 20742 MD USA
}
%----------------------------------------------------------------------------
\abstract{%
We review recent attempts at dealing with the sign problem in Monte Carlo calculations by deforming the region of integration in the path integral from real  to complex fields. We discuss the theoretical foundations, the algorithmic issues and present some results for low dimensional field theories in both imaginary and real time.
}
%----------------------------------------------------------------------------
\maketitle
%----------------------------------------------------------------------------
\section{Introduction}\label{intro}

The Monte Carlo approach to field theories is based on the path integral representation of the expectation of operators given by the path integral:
\beq\label{eq:path_minkowski}
\langle \mathcal{O}\rangle = \frac
{\int D\phi e^{iS} \mathcal{O}(\phi)}
{\int D\phi e^{iS} },
\eeq where $\phi$ stands for the fields in the theory and $S$ is the action of the theory. After discretization the integral in \eq{path_minkowski} becomes a finite dimensional integral. The numerical computation of an integral of this type is nearly impossible due to the highly oscillatory integrand. The standard procedure to deal with this problem is to use a Wick rotation of the time coordinate $t\rightarrow it$ which transforms the integral into
\beq\label{eq:path_euclidean}
\langle \mathcal{O}\rangle = \frac
{D\phi e^{-S} \mathcal{O}(\phi)}
{\int D\phi e^{-S} },
\eeq where now $S$ stands for the euclidean action. In many cases of interest, the euclidean action $S$ is a real quantity. In this case the integral in \eq{path_euclidean} has the form of a classical thermodynamic average and it can be evaluated by Monte Carlo methods using  {\it importance sampling}. $\langle \mathcal{O}\rangle$ is estimated as 
\beq
\langle \mathcal{O}\rangle \approx \frac{1}{\mathcal{N}} \sum_{a=1}^\mathcal{N} \mathcal{O(\phi^{(a)})},
\eeq where  $\phi^{(a)}$, $a=1,\cdots, \mathcal{N}$ are a set of field configurations distributes according to the probability distribution $P[\phi] \sim e^{S[\phi]}$. However, there are many other  cases of physical interest in which the euclidean action is  not real. In that case, the Monte Carlo method can still be used if one separates  the real and imaginary parts of the action as $S=S_R+i S_I$ and considering the phase $e^{-iS_I}$ as part of the observable with the help of the equation:
\beq
\langle \mathcal{O}\rangle = \frac
{\int D\phi e^{-S} \mathcal{O}}
{\int D\phi e^{-S} }
=
\frac
{\int D\phi e^{-S_R} \mathcal{O} e^{-iS_I} }
{\int D\phi e^{-S_R} }
\frac
{\int D\phi e^{-S_R} }
{\int D\phi e^{-S_R} e^{-iS_I}}
=
\frac{\langle \mathcal{O}e^{-iS_I} \rangle_{S_R}}{\langle e^{-iS_I} \rangle_{S_R}}.
\eeq This method (``reweigthing")  works to the extend that the average phase $\langle e^{-iS_I} \rangle_{S_R}$ is not small and the expectation value does not result from a ratio of very small numbers. In the thermodynamic limit (and at low temperature), where the spacetime volume $\beta V$ diverges, the fluctuations of $S_I$ also diverges and the average sign decreases exponentially  as $\beta V$ grows. This is the sign problem.

Physical systems with sign problems  include some of the most interesting physics problems. Systems at finite density typically suffer from the sign problem. That includes the study of cold, dense hadronic/quark matter, a central problem in nuclear physics and astrophysics. Many models used in condensed matter models, like  the Hubbard model away from half-filling, also have a sign problem. In addition, the direct evaluation of time-dependent quantities, even in thermal equilibrium, based on real time integrals like in \eq{path_minkowski}, lead to integrals like the one in \eq{path_minkowski}
and
 have a particularly nasty sign problem \footnote{In thermal equilibrium, real time correlators can be obtained form an analytic continuation of imaginary time correlators obtained form integrals like \eq{path_euclidean}. The analytic continuation of noise numerical data is, however, exponentially hard due to numerical instabilities.}. Given the importance of all these physical systems it is no surprise that many ideas have been put forward to solve or ameliorate the sign problem, too many to even mention (some reviews can be found in \cite{Schmidt:2006us,Ejiri:2008nv,Karsch:2009zz,deForcrand:2010ys}). There is even an argument showing that the sign problem is a NP-complete problem in at least one specific model, implying that a general solution to the sign problem is extremely unlikely to be found \cite{Troyer:2004ge}.

For one-dimensional integrals, there is a well know way of dealing with the oscillatory integrals leading to the sign problem: the stationary phase of steepest descent method. In the stationary phase method one changes the contour of integration
in the integral $\int dz\ e^{-S}$
from the real axis to a contour on the real plane going through one (or perhaps more) critical points (defined by the condition $S'(z)=0$) such that the phase of the integrand is constant. Since the change of contour, under some circumstances, does not alter the value of the integral, this change of contour provides a way of computing the original integral without having to deal with oscillatory phases. The stationary phase contour has one extra feature: it goes along the direction where the absolute value of the integrand decreases the fastest, hence the alternative name ``steepest descent" for this contour. This makes the stationary phase/steepest descent contour also ideal for semi-classsical evaluations of the integral and it is in this context that it is better known among physicists. For the resolution of the sign problem however, it is the constant phase of the integrand on this contour that is relevant.

In this proceeding we will discuss a recent  approach to the solution to the sign problem that can be viewed as an extension of the stationary phase/steepest descent contour method, but the method {\it does not involve any semi-classical expansion}.  It is based on 
deforming the manifold of integration from real variables $\mathbb{R}^N$ to another N-dimensional manifold embedded on the complexified field space $\mathbb{C}^N$. There are at this point in time a variety of choices for the precise deformation of the manifold of integration to be used. Even larger is the choice of algorithmic ideas suggest to accomplish the task in practice. We will discuss some mathematical generalities related to the change of manifolds of integration in section \ref{sec:math}, one algorithm (and its variations) to compute the path integral in practice in section \ref{sec:generalized} and present two examples of the method in action in section \ref{sec:models}.  In the final section we briefly comment on newer ideas and future prospects.

%----------------------------------------------------------------------------
\section{Complexified fields, thimbles, anti-holomorphic flow}\label{sec:math}

Consider  N-dimensional  integrals of the form encountered in lattice field theory:
\beq
\langle \mathcal{O}\rangle=\frac{
\int d\phi_i\ e^{-S[\phi]} \mathcal{O[\phi]}
}
{
\int d\phi_i\ e^{-S[\phi]} 
}.
\eeq By a generalization of the Cauchy theorem familiar from complex analysis, the same integrals  can be obtained by integrating over an N-dimensional submanifold $\mathcal{M}$ of $\mathbb{C}^N$ parametrized by $\phi_i = \phi_i(\zeta_j)$, $i,j=1,\cdots N$:
\beq
Z = \int_{\mathbb{R}^N} d\phi_i\ e^{-S[\phi]} \mathcal{O[\phi]}
=\int_{\mathcal{M}} d\phi_i\ e^{-S[\phi]} \mathcal{O[\phi]}
=\int d\zeta_i\    {\rm det} \left(\frac{\partial \phi_i}{\partial\zeta_j} \right)   e^{-S[\phi(\zeta)]},
\eeq where $\phi_i$ is now a complex variable, $\zeta_j$ are the real variables parametrizing the manifold $\mathcal{M}$, $S[\phi(\zeta)]$ is the analytic continuation of the action and ${\rm det}J= {\rm det} \left(\frac{\partial \phi_i}{\partial\zeta_j} \right) $ is the jacobian related to the change of variables from $\phi_i$ to $\zeta_j$. Notice that $\phi_i$ are complex and so is the jacobian $J$.

\begin{figure}[t]\label{fig:1D} % no figure before 1st section
  \centering
  \includegraphics[width=7cm,angle=90,clip]{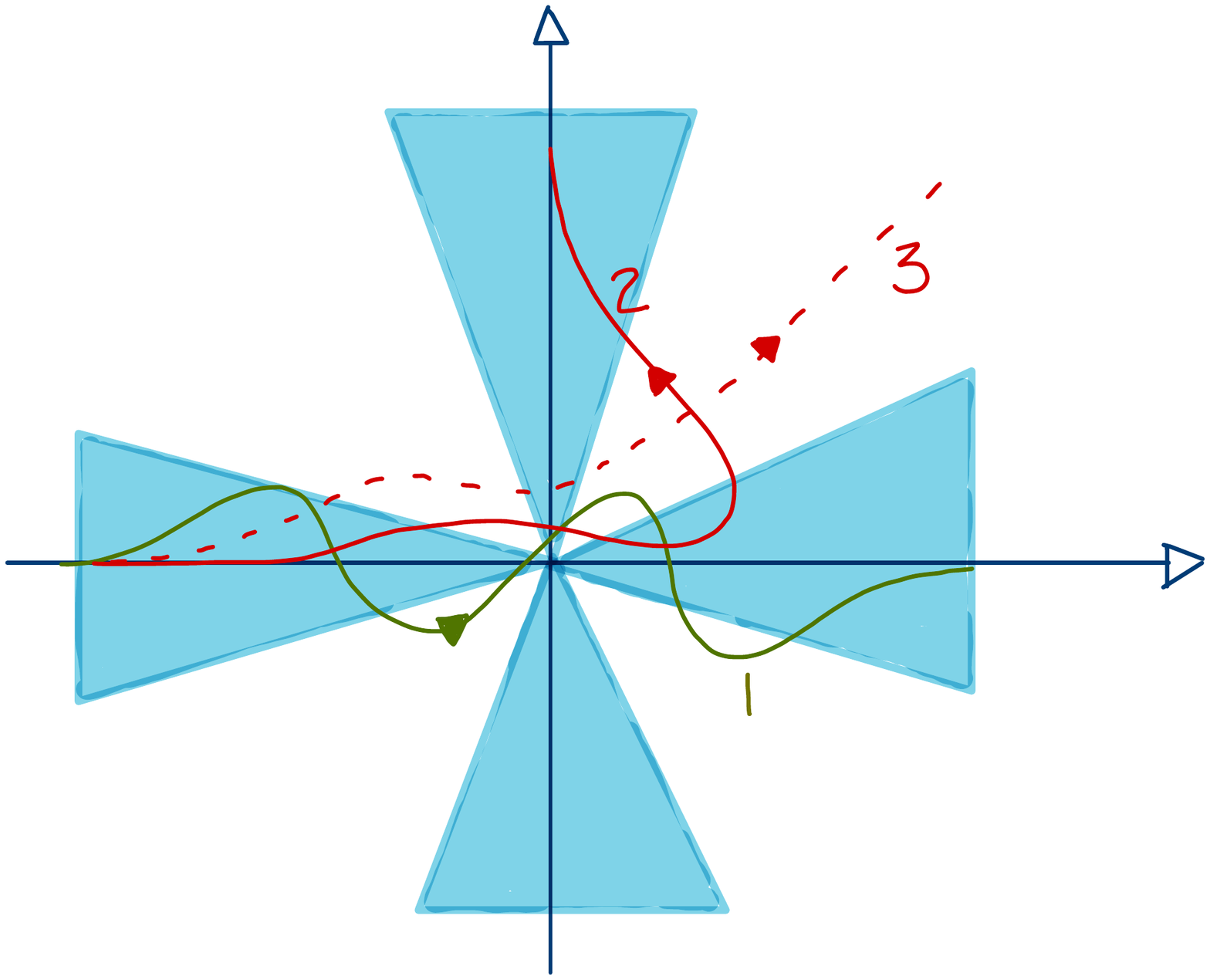}
  \caption{The integral $\int dz\ e^{-z^4}$ is only well defined for contours approaching the infinity along the shaded (blue) areas. The integral over contour $1$ is equivalent to the integral over $\mathbb{R}$. Contour $2$ leads to a different result while the integral over contour $3$ is divergent and not well defined. A continuous deformation of contours from $\mathbb{R}$ to contour $2$ necessarily passes through contours where the integral diverges. This general pattern remains true in higher dimensions. }
  \label{ffig:contours}% Give a unique label
\end{figure}

The change of domain of integration from $\mathbb{R}^N$ to $\mathcal{M}$ does not necessarily lead to the same value of the integrals and care must be taken when performing this step. In the theory of complex functions of one variable the rule  determining which deformations do not change the value of the integral are well known: the deformations allowed are the ones that can be made continuously from the initial to the final contour without crossing any singularity (pole, branch cut) of the integrand including ``poles at infinity". When poles are crossed  during the deformation the value of the integrand jumps discontinuously.
For contours extending towards the infinity the integral is not even necessarily well defined.
 In fact, the integral exists only for contours that approach the point at infinity along certain directions. For the other directions, the integrand does not decay to zero fast enough and the integral diverges. This can be exemplified with the help of the integral $\int_{-\infty}^\infty dz\ e^{-z^4}$. This integral only exists for contours beginning and ending along the shaded regions on fig.\ref{fig:1D}. Since the integrand has no singularity at finite values of $z$ we are free to move the contour around as long as the its beginning and end of the contour do not jump from one of the shaded regions to another. Doing that amounts to crossing the singularity at $z=\infty$.  Thus, the real line is equivalent to the contour labelled $1$ but differs from the one labelled $2$. The integral is not even defined on the contour $3$. Notice that in order to deform contour $1$ into contour $2$, the end of the contour has to necessarily pass through a region where the integrral diverges.
 Two contour are said to be equivalent if they lead to the same value of the integral. The set of all contours (in which the integral is well defined) can thus be divided up in equivalent classes (homology classes) dependent only on which of the shaded regions the initial and final points of the contour lie. 
 
The integrands encountered in lattice field theory do not typically have any singularities. For the most part they are exponential of polynomials (the Boltzmann factor $e^{-S}$ ), perhaps multiplied by another polynomial (observables and/or fermion determinants) \footnote{The formal integration over fermion fields lead to the appearance of propagators computed on the background of bosonic fields. A little thought shows that the poles are illusory and are always cancelled by the fermion determinant (I think this observation to Scott Lawrence).}.  In theories where the field variables are compact (before complexification), the absence of singularities on the integrand is enough to guarantee that any continuous deformation of the domain of integration does not alter the value of the integral. In theories where the fields take values on a non-compact region, like $\mathbb{R}^N$ there may be singularities at asymptotically large values of the fields. Just like one-dimensional integrals, the integrand is well defined only if the asymptotic values are approached from certain ``good" directions. And like one dimensinonal integrals, the set of ``good" directions are grouped in discrete sets (homology classes) and the integral takes on a different value in each of these groups \cite{fedoryuk,pham,kaminski,Witten:2010zr,Witten:2010cx}. As the domain of integration is changed in such a way to make the asymptotic regions go from  one ``good direction" to another, the value of the integral jumps discontinuously (a ``pole at the infinity" was crossed). In between two ``good regions" the integral diverges. Thus, a sufficient condition for a deformation not to change the value of the integral is that it can be continuously connected with the initial domain of integration and that, at all intermediate steps of the deformation, the integral is well defined. This conditions guarantee that the domains lie in the same homology class and can be used as a integration domain instead of the original $\mathbb{R}^N$ .

%%%%%%% 1 D contours 

%%%%%%%%%%%%%%%%%%%

Now that we know that the domain of integration can be moved to a different one in complex space the question is whether this is advantageous. For instance, we may look for a $N-$dimensional manifold  $\mathcal{M}$ embedded in $\mathbb{C}^N$ on which the phase $e^{-i S_I}$ is constant. In fact, the condition $S_I = {\rm constant}$ is one single condition on a function of $2N$ real variables,  defining a $2N-1$ submanifold of $\mathbb{C}^N = \mathbb{R}^{2N}$;  there are certainly many $N$ dimensional manifolds in that $2n-1$ dimensional subspace.
But not all manifolds $\mathcal{M}$ where $e^{-i S_I}$ is constant is suitable.
 For instance,  if the Boltzmann factor $e^{-S_R}$ defines a complicated multimodal landscape on $\mathcal{M}$ such a manifold will be unsuitable for Monte Carlo calculations despite the absence of a sign problem. 
We will first show, as a proof of principle, that manifolds with constant phase and well behaved probability landscapes exist. This construction can be thought out as a multidimensional generalization of the steepest descent/stationary phase contour of one-dimensional integrals.

\subsection{Anti-holomorphic flow and thimbles}\label{thimbles}

Consider the {\it anti-holomorphic} flow defined by the equations\footnote{referred to here on as ``the flow".}
\beq\label{eq:flow}
\frac{d\phi_i}{dt} = \overline{ \frac{\partial S}{\partial\phi_i} },
\eeq where the bar denotes complex conjugation. The important property of the flow is that it increases the real part of the action while keeping the imaginary part constant. In fact,
\beq
\frac{dS}{dt} = \frac{\partial S}{\partial \phi_i} \frac{d\phi_i}{dt}  = 
\frac{\partial S}{\partial \phi_i} \overline{ \frac{\partial S}{\partial \phi_i} }
=  \left| \frac{\partial S}{\partial \phi_i} \right|^2 \geq 0.
\eeq This property can also be seen by splitting the flow in its real and imaginary part we can see that it is the gradient flow for the real part of the action $S_R$ and the hamiltonian flow for the ``hamiltonian" $S_I$. The flow has some critical (fixed) points $\phi^0$ where $\partial S/\partial \phi_i =0$, which we will assume to be isolated. We can understand the behavior of the flow around a critical point by looking at its linearized form:
\beq\label{eq:flow_J}
\frac{d \Delta\phi_i}{dt} = \overline{ H_{ij} }  \overline{ \Delta\phi_j  },
\eeq where $\Delta\phi=\phi-\phi^0$ is the displacement from the critical point and $H_{ij} =  \frac{\partial^2 S}{\partial \phi_i\partial\phi_j}|_{\phi=\phi^0} $ is the hessian. The linearized equations have the solutions

\beq\label{eq:linear_flow}
\phi_i(t) = \phi^0_i +\sum_{a=1}^{2N} e^{\lambda^a t}  \mathbf{e}^a_i ,
\eeq where the vectors $\mathbf{e}^a$ and its ``eigenvalues" satisfy:
\beq\label{eq:takagi}
H_{ij}\mathbf{e}_j^a = \lambda^a \overline{ \mathbf{e}_i^a },
\eeq with real $\lambda^a$ (other solution exist with complex $\lambda^{a}$). 
It can be shown that 
\begin{itemize}
\item $\lambda^a, \mathbf{e}^a$ come in pairs: $(\lambda^a, \mathbf{e}^a)$ and $(-\lambda^a, i\mathbf{e}^a)$ . We will assume that critical point are non-degenerate, that is, all eigenvalues of $H$ are non zero.
\item $\mathbf{e}^a$ form an orthogonal set: $\frac{1}{2}(\overline{\mathbf{e}^a_i} \mathbf{e}^b_i+ \mathbf{e}^a_i \overline{\mathbf{e}^b_i})=\delta^{ab}$.
\item any complex vector in $\mathbb{C}^N$ can be written as a real linear combination of the $\mathbf{e}^a$ as $\mathbf{v} = \sum_{a=1}^{2N} A_a \mathbf{e}^a$ or as a complex linear combination of the $\mathbf{e}^a$ with positive eigenvalues as $\mathbf{v} = \sum_{a=1}^N (A_a+i B_a) \mathbf{e}^a$.
\end{itemize} 

From \eq{linear_flow} we see that the flow around a critical point is repulsive along $N$ of the $2N$ (real) directions, the ones given by the eigenvectors $\mathbf{e}^a$ with positive $\lambda^a$.  The flow is attractive along the other $N$ directions. The points reached by the flow along the repulsive directions form the {\it thimble} of the critical point.  It follows from the properties of the flow that the imaginary part of the action is constant on the thimble and that the real part of the action grows monotonically as the flow moves away from the critical point. As such, thimbles are the proper generalization of the steepest descent/stationary phase contour to the many dimensional case.

In general, there are many critical points and corresponding thimbles and the original integral over real fields equals to the integral over a certain combination of thimbles.  There is a recipe to determine whether a particular thimble is part of this combination: a particular thimble is part of the combination of thimbles equivalent to $\mathbb{R}^N$ is,  by evolution of that  thimble by the {\it reverse} flow intersects the original integration domain ($\mathbb{R}^N$). An intuitive argument for this statement will be presented on the next subsection.

The idea of using thimbles as the integration region in order to solve the sign problem in path integrals was put forward in \cite{Cristoforetti:2012su}. There are two main problems in transforming this idea into a practical method: 
\begin{enumerate}
\item There is no local characterization of the points belonging to a thimble so there is no obvious, computationally cheap way of making Monte Carlo proposals that lie on the thimble.
\item Except for the simplest toy models it is impossible to find all critical points, their thimbles and to determine which thimbles contribute to the original integral.
\end{enumerate}

Problem number 1 has been addressed with several algorithmic ideas. One is the solution of the Langevin equation on the curved thimble, with an appropriate way of projecting the noise term on the tangent plane to the thimble \cite{Cristoforetti:2013wha}. The projection is the computationally most expensive part of the algorithm. A second method is an adaptation of the hybrid Monte Carlo algorithm to sampling on a thimble \cite{Fujii:2013sra}. Here again the difficult and expensive part is the projection  that guarantees the Monte Carlo chain stays on the thimble. The third method, which we will describe in more detail on the next section, uses a map of the thimble (in fact, all the relevant thimbles) by points on $\mathbb{R}^N$. An effective action on the space of real fields is then used with some standard method, like the Metropolis-Hastings algorithm. This last approach has the advantage of being able, in principle, of computing the integral over {\bf} all the relevant thimbles at once, even if the existence and location of these thimbles is not known a priori.

Regarding the problem 2 above, it has been argued \cite{Cristoforetti:2012su} using universality arguments that the integral over a single thimble is likely to agree with the full result in the continuum limit. Also, the contribution of every thimble is suppressed by a Boltzmann factor $e^{-S^0_R}$, where $S^0$ is the  the action at the critical point, and one can imagine that at weak coupling or for large spacetime volumes the integral will be dominated by a single thimble. It is, however,  very hard to put any of these arguments on a firm footing. The contribution of a thimble depends not only on $e^{-S_R}$ but also on its ``entropy". As the spacetime volume increases or the lattice spacing decreases the position and number of thimbles changes. Finally, there is a phase $e^{-i S^0_I}$ multiplying the contribution of each thimble; the contribution of several thimbles may cancel or, if many thimbles contribute significantly, a sign problem might reappear when combining their contributions. The plausibility of the one thimble dominance can also be gauged against some of its consequences. In QCD, for instance, an instanton configuration is a solution of the classical equations of motion and so there is a thimble associated with it. If the thimble associated with the trivial configuration dominates the partition function, the instanton contributions and its physical consequences have to disappear in the infinite volume/continuum limits. Fortunately, there are ways to proceed without making any assumptions about the dominance of one or more thimbles and, in fact, those arguments can be tested with numerical calculations. The few examples where the thimble structure was fully understood do not have a continuum or thermodynamics limit where these questions ca be addressed
\cite{Fujii:2015bua,Tanizaki:2015rda,Tanizaki:2014tua,Kanazawa:2014qma,Tanizaki:2016xcu}.

\subsection{Flowing towards the right thimbles}\label{us}

%%%%%%% flowing to thimble 

\begin{figure}[t] % no figure before 1st section
  \centering
  \includegraphics[width=7cm,angle=90,clip]{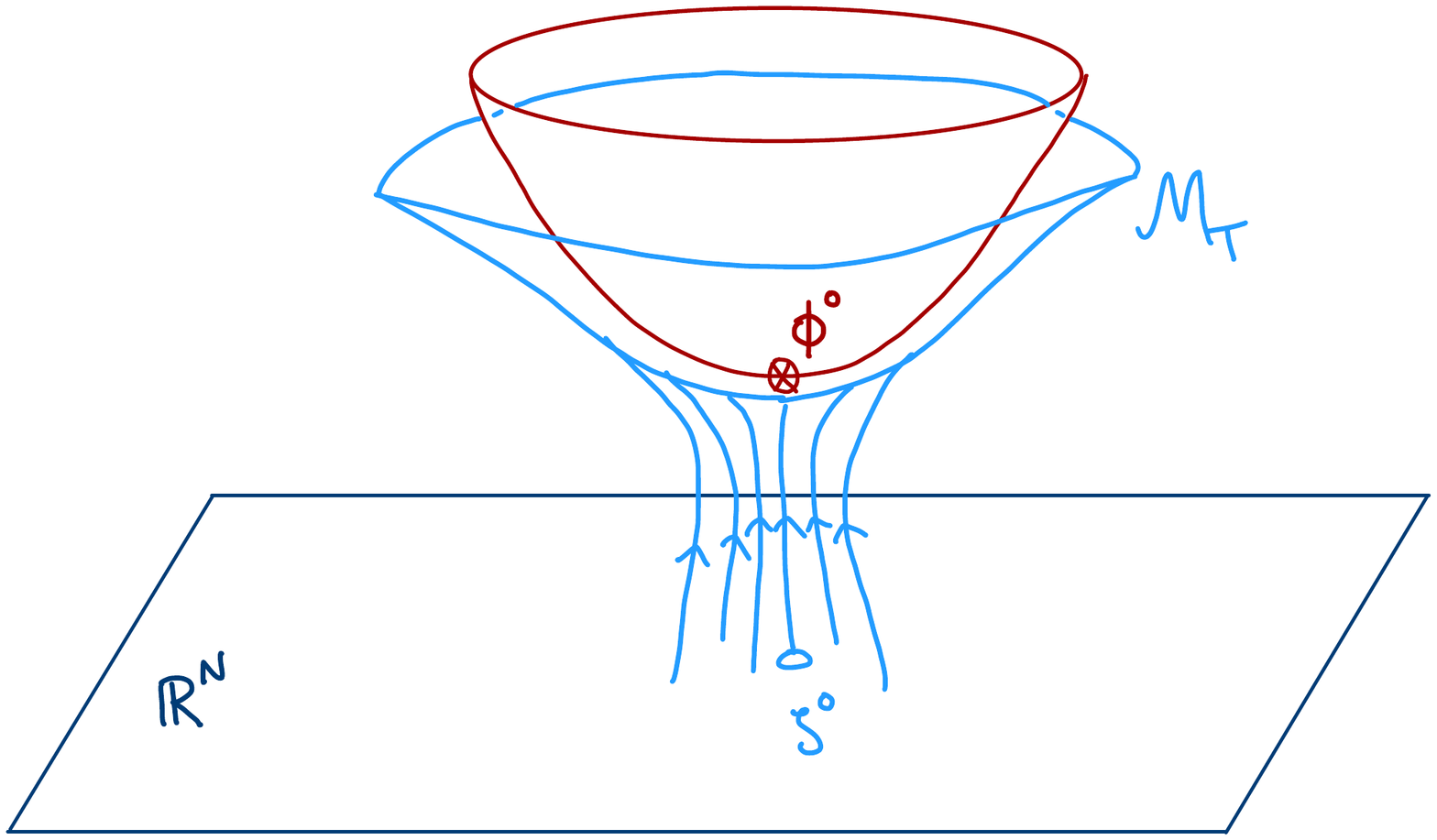}
  \caption{The flow takes point in $\mathbb{R}^N$ to points in $\mathcal{M}_T$. The integral over $\mathcal{M}_T$ equals the one over $\mathbb{R}^N$ at all values of $T$. As $T$ increases,  $\mathcal{M}_T$ approaches the combination of thimbles equivalent to the original domain of integration.}
  \label{fig:thimble}% Give a unique label
\end{figure}
%%%%%%%%%%%%%%%%%%%

Consider taking every point $\zeta$ of $\mathbb{R}^N$ and evolving them by the flow in \eq{flow} by a fixed time $T$. 
The end points $\phi_T(\zeta)$ of the trajectories form a $N-$ dimensional manifold we call $\mathcal{M}_T$.
 Assuming the integrand has no singularities and the flow by time $T$ is well defined for all initial conditions with finite values of the fields, the multidimensional Cauchy theorem applies and the equality of the integrals hinges on the singularities ``at the infinity", that is, whether the two manifolds are in the same homology class. In many  field theories of physical interest, like gauge theories and $\sigma-$models, the field variables are compact, the original domain of integration is not $\mathbb{R}^N$ but some other $N-$ dimensional compact manifold $\mathcal{M}_0$ and there are not asymptotic directions where the integral might diverge. The integral over $\mathcal{M}_T$ thus equals the one over $\mathcal{M}_0$. In the case of non-compact field variables we note that the flow monotonically increases the real part of the action and, consequently, decreases the weight $e^{-S_R}$. Since the integral was well defined over $\mathbb{R}^N$ and the integrand only decreases as $T$ increases, it is reasonable to assume that the integral remains finite and well defined for all intermediate values of $T$. Since the value of the integral can only change, as the manifold on integration changes, after going through a region where the integral diverges (at large values of the field), we conclude that the integral will not change in the non-compact case either. 

We can now find out what happens to the manifold $\mathcal{M}_T$ as $T$ is increased. 
One point $\zeta^0$ will approach the critical point $\phi^0$ asymptotically (if there isn't such a $\zeta^0$ that particular thimble is not part of the set contributing to the original integral). The surrounding points will flow towards points with larger and larger values of $S_R$. However, no point can cross a thimble since the flow is tangent to them. They can only asymptotically approach the thimbles. 
As a result, the flow ``automatically" maps the real fields into a manifold that is close to the right combination of thimbles equivalent to the integration over $\mathbb{R}^N$. As $T$ is increased, $\mathcal{M}_T$  becomes closer to the thimbles.
This situation is pictured in fig.\ref{fig:thimble}.

There is a jacobian involved in the parametrization of $\mathcal{M}_T$ in terms of the real fields (the initial conditions of the flow). If the result of flowing the point $\zeta_i$ by a time $T$ is the (complex) point $\phi_i(\zeta)$, the jacobian $J={\rm det}( \partial \phi_i(\zeta)/\partial \zeta_j)$ can be computed by taking a basis $( \mathbf{v}^{(a)} )$ of $\mathbb{R}^N$ and evolving each of its vectors by the flow
\beq
\frac{d v_i^{(a)} }  {dt} = \overline{ \frac{\partial^2 S}{\partial z_i \partial z_k}  }  \overline{  v_k^{(a)}  }.
\eeq The computation of the jacobian is extremely expensive. Not only $N$ sets of $N$ computed differential equations need to be solved but the  determinant of a $N \times N$ dense complex matrix need to be computed, leading to a computational cost of order $N^3$\footnote{A method for the computation of the residual phase is discussed in \cite{Cristoforetti:2014gsa}}. Also, the jacobian is, in general, a complex number: $J = |J| e^{i\alpha}$. A Monte Carlo evaluation of an integral over $\mathcal{M}_T$ requires then the reweighting of the ``residual phase" $\alpha$ as well as $e^{-i S_I}$:

\beq\label{eq:reweigth}
\langle \mathcal{O}\rangle = 
\frac
{\int _{{\mathcal{M}_T}}D\phi e^{-S} \mathcal{O} }
{\int_{{\mathcal{M}_T}} D\phi e^{-S} }
=
\frac
{\int D\phi e^{-S_R+\log|J|} \mathcal{O} e^{-iS_I+i\alpha} }
{\int D\phi e^{-S_R+\log|J|} }
\frac
{\int D\phi e^{-S_R+\log|J|} }
{\int D\phi e^{-S_R+\log|J|} e^{-iS_I+i\alpha}}
=
\frac{\langle \mathcal{O}e^{-iS_I+i\alpha} \rangle_{S_R}}{\langle e^{-iS_I+i\alpha} \rangle_{S_R}}.
\eeq At large values of $T$, the manifold $\mathcal{M}_T$ is  close to the (correct set of) thimbles and, since $S_I$ is a constant over each thimble we don't expect $S_I$ to fluctuate much unless many thimbles with different phases contribute almost equally to the integral. Experience shows also that for the variety of models and parameter space studied up to now the residual phase also fluctuates little and there is no practical problem in reweighting it. 

For weakly coupled theories (not necessarily perturbative) this is expected. In fact, free theories,  which have a quadratic action, have flat thimbles with constant $\alpha$. The experience accumulated investigating several models over a large swatch of parameters space suggests that the fluctuation of $\alpha$ and, for large enough flow time $T$, the fluctuations of $S_I$ also, are small and easily reweigthed. Of course, this always has to be verified in a particular calculation. An important theoretical question is whether this experience will extend to the continuum and infinite volume limits in 4 dimensional theories.

\section{The generalized thimble method}\label{sec:generalized}

\subsection{Basic algorithm}\label{sec:basic}

The observations made in the previous sections suggest a way of computing the integral over 
a manifold $\mathcal{M}_T$ in the same homology class as $\mathbb{R}^N$. This set of manifolds includes, if one takes a large enough flowing time $T$, the correct combination of thimbles equivalent to the real field space. The idea is to parametrize each point $\phi$ in $\mathcal{M}_T$ by the point $\zeta$ in $\mathbb{R}^N$ that flows to it: $\phi=\phi_T(\zeta)$, where $\phi_T(\zeta)$ is the map given by flowing $\zeta$ by a time $T$. This way the action on $ \mathcal{M}_T$ can be thought of as a function of the real field $\zeta$,  $S=S[\phi_T(\zeta)]$. The jacobian can also be written in terms of the real variables $\zeta$ and the simulation over $\mathcal{M}_T$ is equivalent to a simulation over real fields $\zeta$ with an effective action equal to $S_{eff}= S[z(\zeta)] -\log{\rm det} J(\zeta)$. Since $S_{eff}$ is complex, its  imaginary part needs to be reweighted, as shown in \eq{reweigth}. The sampling of configurations $\zeta_i$ with distribution $P[\zeta] \sim e^{-\mathbb{R}e (S_{eff} )} $ can be done with any of a number of methods like, for instance, the Metropolis algorithm. To summarize, the algorithm we propose is \cite{Alexandru:2015xva,Alexandru:2015sua}:

\begin{enumerate}\label{eq:algo}
\item Start from a initial point $\zeta$ in $\mathbb{R}^N$ and the corresponding effective action $S[\phi_T(\zeta)]-\log{\rm det}J(\zeta)$ where $\phi_T(\zeta)$ is obtained by integrating the flow equations \eq{flow} with $\zeta$ as the initial conditions and solving \eq{flow_J}.
\item  Make a random proposal $\zeta'=\zeta+\Delta\zeta$ in $\mathbb{R}^N$ and compute the corresponding point $\phi'=\phi_T(\zeta')$ and comoute the action $S[\phi']$, the jacobian $J(\zeta)$ and the effective action $S[\phi']-\log{\rm det}J(\zeta')$.
\item Accept or reject the new configuration $\zeta'$ based on the real part of difference of effective actions between points $\zeta$ and $\zeta'$ with the usual Metropolis probabilities.
\item Return to 2 until a suitably large number of (statistically independent) configurations is collected.
\item Compute the average of the phase $e^{-iS_I+i\alpha}$ and the observable $\mathcal{O}e^{-iS_I+\alpha}$ over the collected configurations using the imaginary parts of the action ($\mathbb{I}m S[\phi]$) and the jacobian ($\alpha=\mathbb{I}m\log{\rm det}J$). 
\end{enumerate} An important observation is that any value of the flowing time $T$ leads to a manifold $\mathcal{M}_T$ equivalent to $\mathbb{R}^N$. If $T$ is too small, however, the phase $e^{-iS_I}$ will fluctuate a lot and the uncertainty on the ratio $\langle\mathcal{O} e^{-iS_I+i\alpha} \rangle/\langle e^{-iS_I+i\alpha}\rangle$ will make the method unpractical. In this case a larger value of $T$ is required. There is, however, a problem that can arise at too large a value of $T$. At large $T$, a small region of $\mathbb{R}^N$ is mapped on to a large region of the thimble. Suppose two different thimbles contribute significantly to the integral. Since every thimble is mapped by one small region of $\mathbb{R}^N$,
so the region  of $\mathbb{R}^N$ contributing significantly to the integral will split up into
 two small, isolated regions. In other words, the  probability distribution $\sim e^{-\mathbb{R}e(S_{eff})}$ is multimodal, which is a situation well known to be challenging for Monte Carlo evaluations.
 This poses a serious difficulty for the integration over $\mathcal{M}_T$ mapped by the flow as done here. This will be exemplified later.

 It would be useful at this point  to list some advantages and disadvantages of this algorithm over previously proposed ones. The main advantage is that no previous knowledge about the position or shape of thimbles, or whether they contribute to the integrals, is necessary. The main disadvantages are i) the high computational cost of the jacobian,  ii) the multimodality of the distribution in parameter space $\zeta$ if $T$ is too large and iii) the fact that isotropic proposals in $\zeta$-space ($\mathbb{R}^N$) result in highly anisotropic proposals in $\mathcal{M}_T$ which do not match the probability distribution profile in $\mathcal{M}_T$ and make the sampling inefficient. This last problem is much reduced in many, but not all models, by rescaling the size of the proposals along different directions according to the expectation given by a quadratic approximation of the action \cite{Alexandru:2015xva}.

Some extensions of the basic algorithm were proposed to deal with the trapping of Monte Carlo chains due to the multimodality of the distributions in parameters space \cite{Fukuma:2017fjq,Alexandru:2017oyw}. Some ways of dealing with the high cost of the jacobian are described in the next subsection.

%%%%%%% flowing to thimble 

\begin{figure}[t] % no figure before 1st section
  \centering
  \includegraphics[width=7cm,angle=90,clip]{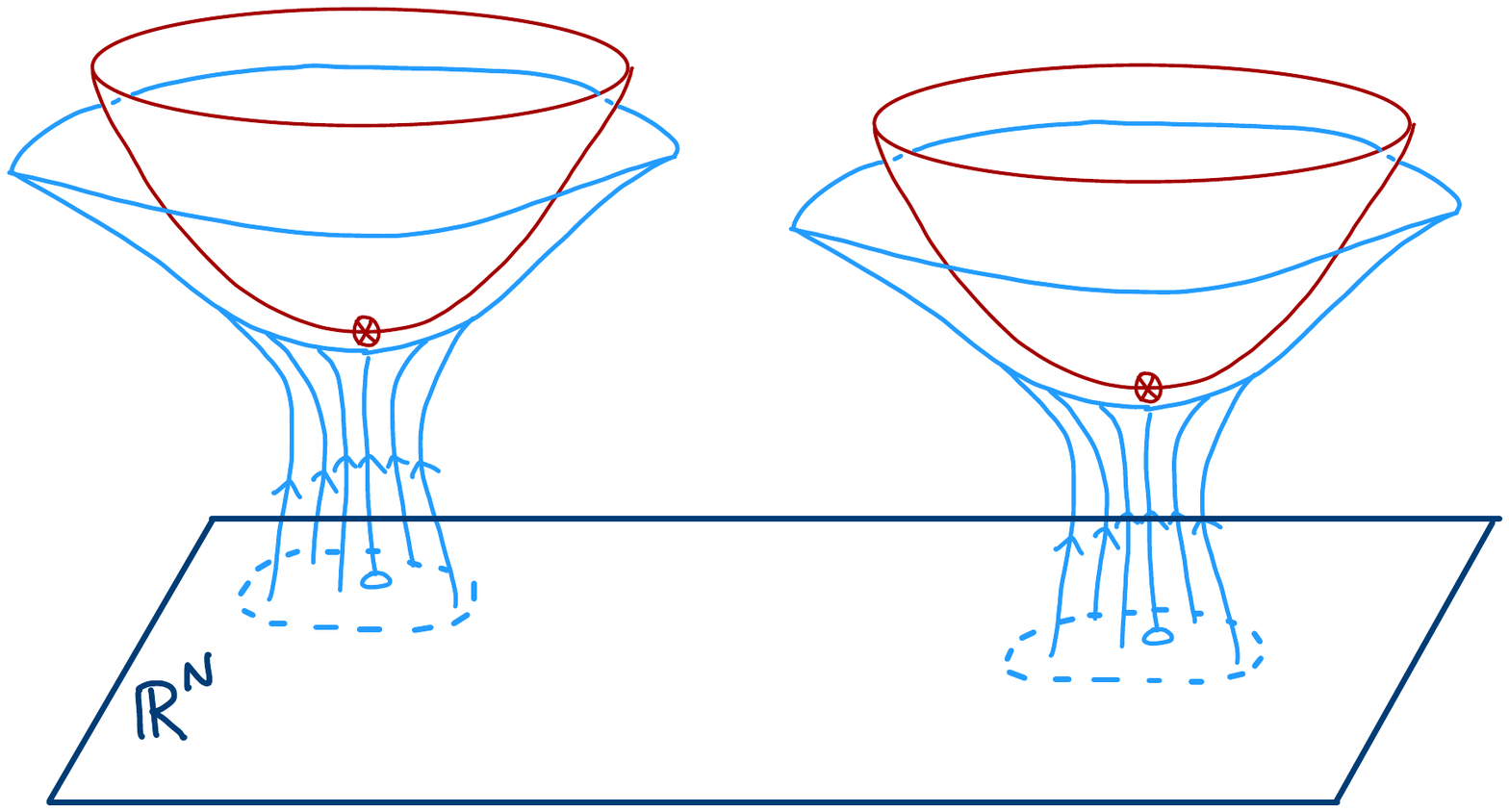}
  \caption{The flow maps point in $\mathbb{R}^N$ to points in $\mathcal{M}_T$. The integral over $\mathcal{M}_T$ equals the one over $\mathbb{R}^N$ at all values of $T$. As $T$ increases,  $\mathcal{M}_T$ approaches the combination of thimbles equivalent to the original domain of integration.}
  \label{fig:thimbles}% Give a unique label
\end{figure}
%%%%%%%%%%%%%%%%%%%

We know that at large enough $T$, the manifold $\mathcal{M}_T$ approaches the correct combination of thimbles where the sign problem is minimized (it doesn't disappear completely because more than one thimble, with different phases $e^{-i S_I}$, may contribute). It is reasonable to expect that at intermediate values of $T$ the sign problem will be somewhat ameliorated. 
Before we delve into algorithmic issues, it would be useful to understand how in practice that occurs.

I may seem a little mysterious how the sign problem can be ameliorated by computing the path integral over  $\mathcal{M}_T$ instead of over $\mathbb{R}^N$. After all, the flow preserves the imaginary part of the action and therefore does not change the fluctuating phase $e^{-i S_I}$ which is the origin of the sign problem. The apparent paradox is resolved when one realizes that only a small region of $\mathbb{R}^N$ is mapped into points near a particular thimble. If $\zeta^0$ is mapped to $\phi_T(\zeta) \approx \phi^0$, ($\phi^0$ being a critical point of the action) a small neighborhood around $\zeta$ is mapped near the thimble attached to $\phi^0$. As $T$ is increased the size of the neighborhood around $\zeta^0$ mapping the same region near the thimmble  shrinks.
The points of $\mathbb{R}^N$ that are not mapped to point near thimbles flow towards infinity or other regions of $\mathbb{C}^N$ where the action diverges (if any exists). The consequence of this picture is that there is little phase ($e^{-i S_I}$) variation on points near the thimble and all the phase variation found in $\mathbb{R}^N$ is pushed towards regions of $\mathcal{M}_T$  with very large values of $S_R$ and, consequently, contribute little to the integral. In the limit $T\rightarrow\infty$ only an infinitesimal region around $\zeta^0$ is mapped on to the thimble. All remaining points in $\mathbb{R}^N$ flow to regions where $S_R$ diverges (usually the infinity) and do not contribute to the integral. This situation is pictured in fig.~\ref{fig:thimbles}.

\subsection{Cheaper jacobians: estimators, Grady-style algorithm}\label{sec:pseudo}
One strategy to deal with the high computational cost of the jacobian is to have a rough estimator of it to be used to generate the Monte Carlo chain. The difference between the estimator and the correct jacobian can be included as a reweigthing factor, as it is done with the phase of the jacobian. We have found two estimators that seem to be useful in (some) of the models we have studied. The first estimator of ${\rm det}J$ is $W_1$ defined by
\beq
\log W_1(\zeta) = \int_0^Tdt\ \sum_{a=1}^N\mathbf{e}_i^{(a)\dagger} \overline{  \frac{\partial^2S(\phi(t))}{\partial\phi_i \partial\phi_j} }  \overline{\mathbf{e}_j^{(a)}}
\eeq where $\mathbf{e}^{(a)} $ is the (complex) basis formed by the solutions to \eq{takagi} with positive eigenvalues. The hessian $ \frac{\partial^2S}{\partial\phi_i \partial\phi_j}$ is to be computed along the flow trajectory starting at $\zeta$ while the basis $\mathbf{e}^{(a)} $ is computed from the hessian at one critical point of a ``dominating" thimble. For quadratic actions (free theories), $W_1=J$ and, consequently, it is to be expected that $W_1$ is a good estimator in weakly coupled theories. Experience shows that it is an useful estimator even in theories which, by any measure, are strongly coupled. The second estimator $W_2$ is even simpler to compute:
\beq
\log W_2 = \int_0^T dt\ {\rm tr} \left( \overline{  \frac{\partial^2S(\phi(t))}{\partial\phi_i \partial\phi_j} } \right)
=
\int_0^Tdt\ \sum_{a=1}^N\mathbf{e}_i^{(a)\dagger} \overline{  \frac{\partial^2S(\phi(t))}{\partial\phi_i \partial\phi_j} }  \mathbf{e}_j^{(a)}.
\eeq $W_2$ is the determinant of the matrix $W_{2\ ij}$satisfying the equation
\beq
\frac{dW_{2\ ij}}{dt} = \overline{  \frac{\partial^2S}{\partial\phi_i \partial\phi_k} } W_{2\ kj},
\eeq with initial condition $W_{2\ ij}(t=0) = \delta_{ij}$ instead of 
\beq
\frac{dJ_{ij}}{dt} = \overline{  \frac{\partial^2S}{\partial\phi_i \partial\phi_k} } \overline{J_{kj}},
\eeq satisfied by $J$. This suggests that $W_2$ will be a good estimator if $J$ is nearly real, which agrees with our experience. We emphasize, however, that the use of these estimators does not introduce an uncontrolled error in the calculation since, when making measurements, the difference between the estimator and the correct jacobian is reweigthed.

A more elegant way of dealing with the cost of the jacobian, which we will call the Grady algorithm, is given by a modification of our algorithm in \eq{algo} that  includes a way of bypassing the computation of the jacobian by adapting
the Grady algorithm \cite{Grady:1985fs,Creutz:1992xwa} originally created to deal with fermion determinants.
  The main idea \cite{Alexandru:2017lqr} is to make a modified proposal that is isotropic in the variable $\phi_T(\zeta)$, not in $\zeta$, and then correct for that by modifying the accept/reject step. The proposal probability is given by
\beq
\mathcal{P}(\zeta\rightarrow \zeta') = e^{-\Delta\zeta J^\dagger J\Delta\zeta} \sqrt{{\rm det }J^\dagger J}
=
e^{-\eta_\parallel^\dagger \eta_\parallel} \sqrt{{\rm det} J^\dagger J},
\eeq where $\Delta\zeta = \zeta' - \zeta$, $J=J(\zeta)$ is the jacobian matrix and $\eta_\parallel$ is a complex vector tangent to $\mathcal{M}_T$. The action of $J$ on a real vector is the same as evolving the vector by the equations \eq{flow_J}. The result is, by construction, a vector tangent to $\mathcal{M}_T$: $J \Delta\zeta = \eta_\parallel$. However, flowing an imaginary vector, normal to $\mathbb{R}^N$, does not give a complex vector normal to $\mathcal{M}_T$. Thus, the proposal with probability above can be found by picking a random vector from the distribution $\sim e^{-\eta^\dagger\eta}$, solving iteratively the equation $J\Delta\zeta = \eta$ and then computing $\eta_\parallel = J \re(\Delta\zeta)$. The action of $J^{-1}$ on a complex vector is obtained by flowing the real and imaginary parts of the vector {\it separately} by the {\it inverse} of the flow in \eq{flow_J}. The backward flow has to be applied to the real and imaginary parts separately as the flow evolution is not a linear operator (it has an anti-linear part). The accept/reject step proceeds as follows. After a proposal $\zeta'$ is chosen, a random real vector $\xi$ is chosen with probability distribution $p[\xi] = e^{-\xi^\dagger J^{'\dagger} J' \xi   J^{'\dagger} J' } \sqrt{{\rm det} J^{'\dagger}J'}$ by iteratively solving $J'\xi = \eta$, where $\eta$ is chosen from a gaussian distribution $\sim e^{-\eta^\dagger \eta}$. The probability of acceptance of $\zeta'$ is given by ${\rm min} (1, e^{-S_{eff}[\zeta']+S_{eff}[\zeta]})$. It is easily shown that this procedure satisfies detailed balance\cite{Alexandru:2017lqr}.

%%%%%%% thirring thimbles 

\begin{figure}[t] % no figure before 1st section
  \centering
  \includegraphics[width=9cm,angle=0,clip]{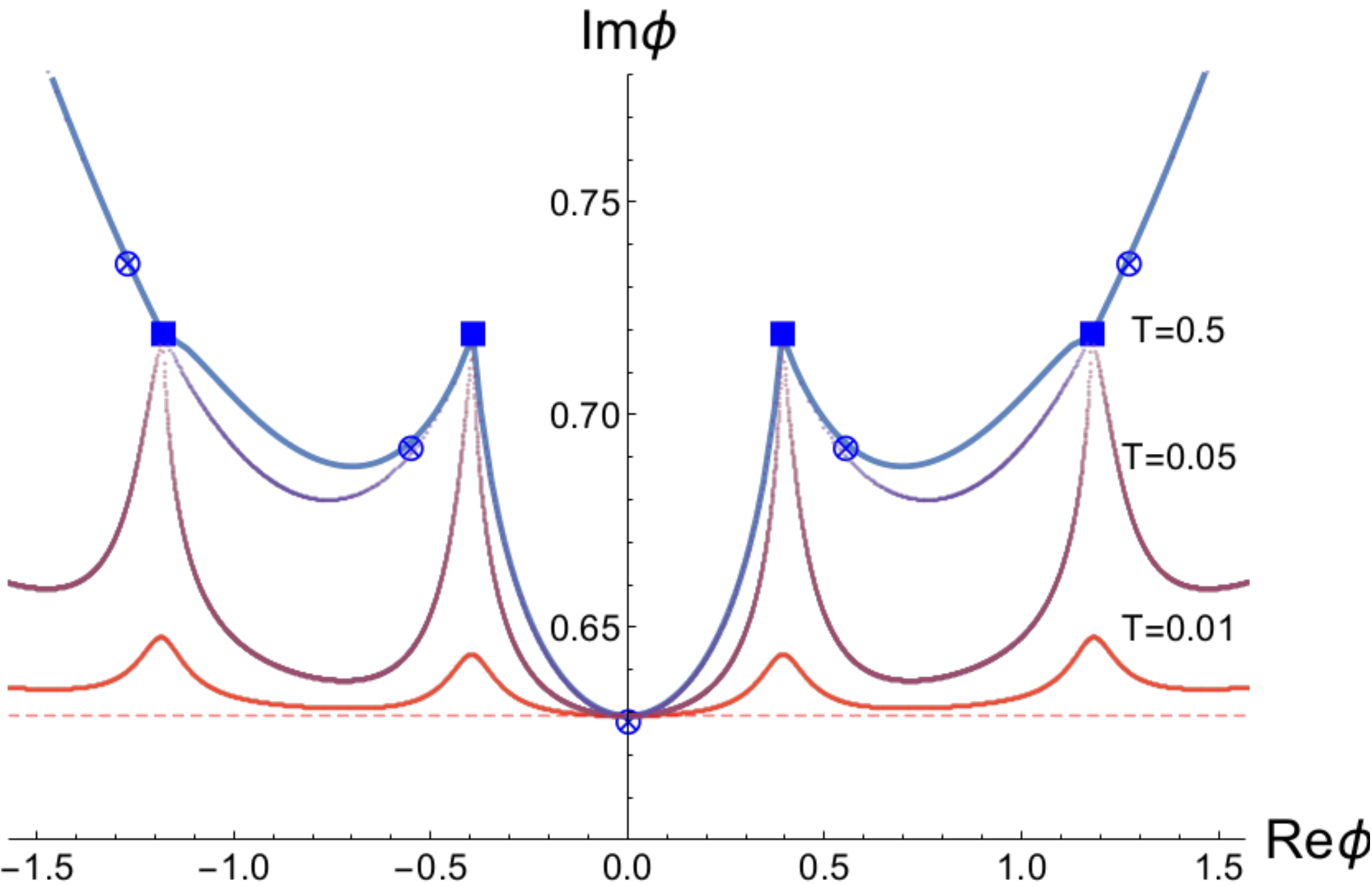}
  \caption{
  Cross-section $A_0(x) = \re\phi + i \im\phi, A_1(x)=0$. The thimbles are shown in blue, the critical points are denoted by a cross and the ${\rm det}D(A)=0$ points are the squares.
  The flow takes point in $\mathbb{R}^N$ to points in $\mathcal{M}_T$. The integral over $\mathcal{M}_T$ equals the one over $\mathbb{R}^N$ at all values of $T$. As $T$ increases,  $\mathcal{M}_T$ approaches the combination of thimbles equivalent to the original domain of integration.}
  \label{fig:thirring_thimbles}% Give a unique label
\end{figure}
%%%%%%%%%%%%%%%%%%%

\section{It all comes together in one example:}\label{sec:models}
\subsection{ The $1+1 D$ Thirring model}\label{sec:thirring}

After quite a bit of theoretical discussion it will be good to see how these ideas apply -- and how well they fare -- in an actual numerical calculations. We choose the $1+1$ dimensional Thirring model at finite density (and temperature) as a suitable testing ground. It has the advantage of having less degrees of freedom than a 4-dimensional theory would, while sharing some features with QCD like asymptotic freedom, a formulation where the sign problem appears (only) at finite density and a complex fermion determinant (bosonic sign problems were studied using similar approaches in \cite{Cristoforetti:2013wha,Mukherjee:2013aga,DiRenzo:2015foa,Alexandru:2016san} ). The model is defined in the continuum by the (euclidean) action

\beq\label{eq:S_continuum}
S=\int d^2x\ [\bar\psi^\alpha (\slashed{\partial}+\mu \gamma_0 +m)\psi^\alpha 
+ \frac{g^2}{2N_F}\bar\psi^\alpha\gamma_\mu\psi^\alpha \bar\psi^\beta\gamma_\mu\psi^\beta],
\eeq 
where the flavor indices take values $\alpha,\beta=1,\ldots,N_F$, $\mu$ is the chemical potential and the Dirac spinors $\bar\psi,\psi$ have two components. Both Wilson and staggered discretizations of the fermions were used in \cite{Alexandru:2016ejd}; we will describe here the results with Wilson fermions. In that case the discretized action chosen is

\beq
\!\!\!\!\!S=\sum_{x,\nu} \frac{N_F}{g^2} (1-\cos A_\nu(x)) 
+
\sum_{x,y} \bar\chi^\alpha(x) D_{xy}(A)  \chi^\alpha(y)\,,
\eeq 
with 
\beq
D_{xy} = 
\delta_{xy} - \kappa \sum_{\nu=0,1}  
\Big[ 
 (1-\gamma_\nu) e^{i A_\nu(x)+\mu \delta_{\nu 0}} \delta_{x+\nu, y}
    + (1+\gamma_\nu) e^{-i A_\nu(x)-\mu \delta_{\nu 0}}  \delta_{x, y+\nu}
\Big],   
\eeq 
 Notice we chose to use auxiliary bosonic variables that are periodic so the original manifold of integration of the path integral is $(S^1)^N$. This choice makes the model close to gauge theories and avoids some of the discussion we had in section \ref{sec:math} regarding the asymptotic behavior of 
$\mathcal{M}_T$.
The integration over the fermion fields leads to
\beq
\!\!\!S=N_F 
\left(  
\frac{1}{g^2}\sum_{x,\nu} (1-\cos A_\nu(x)) - \log\det D(A).
\right),
\eeq This action describes $N_F$ Dirac fermions in the continuum. For $\mu\neq 0$ the determinant $\det D(A)$ is not real so this model cannot be simulated by standard Monte Carlo techniques. From here on we choose $N_F=2$. 
In the continuum this model is solvable but, since we will work not too close to the continuum.
The thimble structure of this model is partially understood, specially in the subspace $A_0(x) = \re\phi + i \im\phi = $ constant, $A_1(x)=0$ where it is similar to the one of the $0+1$ dimensional Thirring model \cite{Pawlowski:2014ada,Fujii:2015bua,Fujii:2015rdd}. There are a number of critical points in the $A_0(x)=\phi, A_1(x)=0$ complex plane. There are also points where the action diverges. They arise because the fermion determinant vanishes for some values of the $A_\mu$ fields. They do no imply in any singularity in the integral, they actually correspond to zeros of the integrand and the contribution of the fields around them is suppressed. Since the action approaches $S_R\rightarrow\infty$ there, they are attractors of the anti-holomorphic flow. The picture that emerges is that these singularities of the action arise at the border of the thimbles and different thimbles approach these singularities from different directions. The manifold defined by ${\rm det} D(A)=0$ has $2N-2$ (real) dimensions. Two manifolds, each with dimensions $N$, generically meet only at isolated points with dimensions 0. But thimbles are tangent to the anti-holomorphic flow that ``seeks" points with divergent action so they end up meeting at a larger dimensional space. This structure is generic to models with fermions and it is depicted on fig.\ref{fig:thirring_thimbles}. 

%%%%%%% thirring results 

\begin{figure}[t] % no figure before 1st section
  \centering
  \includegraphics[width=13cm,angle=0,clip]{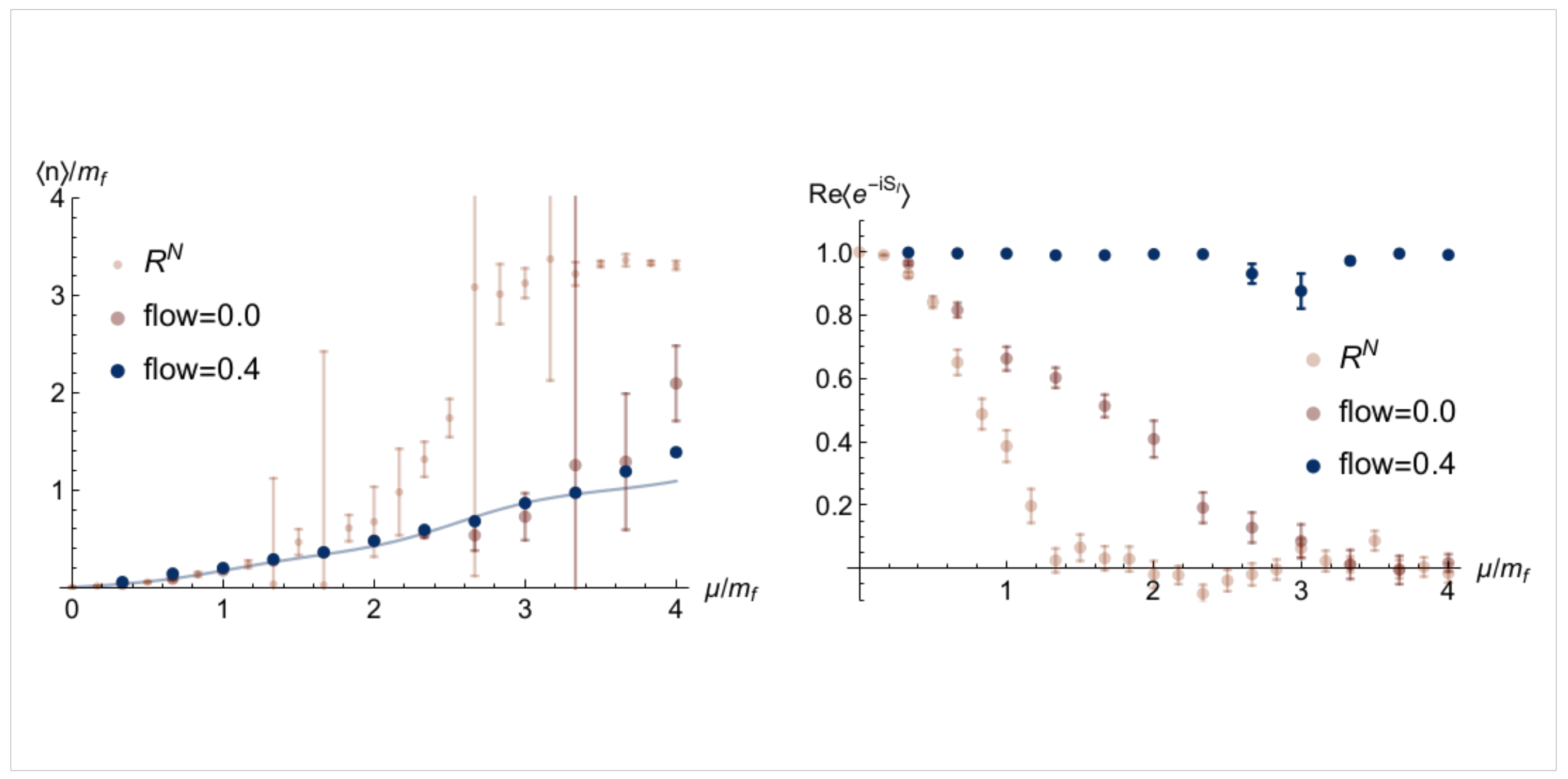}
  \caption{ Fermion density versus chemical potential (left panel) and average sign versus chemical potential (right panel). 
 }
  \label{fig:thirring_results}% Give a unique label
\end{figure}
%%%%%%%%%%%%%%%%%%%

Let us consider some specific values of the parameters for illustrative purposes. With $m=-0.25 $, $g=1.0 $ we find, by performing standard Monte Carlo at zero density on a $10\times 10$ lattice, a fermion mass $am_f = 0.30(1)$ and a bosonic mass (for the state generated in the continuum by $\bar\psi_a \gamma_5 \tau^3_{ab}\psi_b$) of $am_b= 0.44(1)$. Since $m_b << 2 m_f$ these parameters correspond to a strongly coupled model. An attempt at simulating this model with usual techniques, namely, an integration over $(S^1)^N$ reveals that there is a terrible sign problem for all chemical potentials larger $m_f$. In fig.\ref{fig:thirring_results} we show the average sign (right panel) and the fermions density (left panel). It is evident that, even on this small lattice, there is a severa sign problem.

The critical point with the smallest action is the one nearest to $(S^1)^N$ at $\phi=\phi_c=$ purely imaginary and, at least at weak coupling, is expected to dominate the path integral. The tangent plane to the thimble at $\phi_c$ is purely real. If we take the domain of integration to be this tangent plane we expect the sign problem to improve as compared to the one on  $(S^1)^N$ since the action, and therefore its imaginary part, varies only quadratically with the deviation from the critical point. In other words, the tangent plane is a good approximation to the thimble for points near the critical point. Since the the phase $e^{-iS_I}$ is fixed on the thimble, we expect it to vary little on the tangent plane. This expectation is supported by actual calculations shown in fig. \ref{fig:thirring_results}. 
Notice that calculations on the tangent plane are just as computationally cheap as the usual integration over real variables.
Still, for values of the chemical potential high enough ($\mu > 3 m_f$) the sign problem returns, even in these modest lattice sizes. We now use the algorithm described in section \ref{sec:generalized} except that we start the flow not from real values of the field $A_\mu(x)$ but from the $A_0(x)=\phi_c, A_1(x)=0$ plane in order to take advantage of the improvement obtained on the tangent plane. A cross section of the manifold $\mathcal{M}_T$ obtained that way with several flow times  is shown in fig.\ref{fig:thirring_thimbles}. As discussed above, the higher  the value of $T$ the closer $\mathcal{M}_T$ is to the (correct combination of) thimbles. The integrations over each one of these manifolds $\mathcal{M}_T$ give the same result but at different computational cost as the sign problem is much milder at larger $T$. This can be seen on fig.\ref{fig:thirring_results}. The average sign becomes much larger already at $T=0.4$, almost completely solving the sign problem and leading to the results on the left panel of \ref{fig:thirring_results}. There is a price to pay, however, if $T$ is too large. The regions of $\mathcal{M}_T$ contributing significantly to the integral are mapped by regions of  $(S^1)^N$ that are well separated by  barriers of low probability $\sim e^{-S_R}$. This situation is hard to be dealt with by any Monte Carlo calculation (although ``tempered" algorithms have been used to deal with it \cite{Fukuma:2017fjq,Alexandru:2017oyw}). If not extra care is taken (and the computational resources spent), the Monte Carlo chain becomes trapped in one of the modes of the distribution and only the region of $\mathcal{M}_T$ close to one of the thimbles is properly sampled. In cases where the contribution of the other thimbles is sizable (given the statistical uncertainty of the measurements), the final result is erroneous. On the other hand the trapping due to excessive flow can be used to separate the contribution of different thimbles to a given observable. In \cite{Alexandru:2015sua}, the $0+1$ dimensional Thirring model was studied and it was shown that the trapped calculation, sampling only one thimble, gives results that differ by small but statistically significant, amounts from the exact analytic result (the single thimble calculation discussed in \cite{Fujii:2015vha} also agrees with our trapped calculation.). The approach to the thermodynamics and continuum limits in this model does not seem to bring any extra difficulties in this model \cite{Alexandru:2015sua}. In particular, the plateau structure on observables due to shell effects in a finite volume are very visiby and correctly described by the calculations. As pointed out in \cite{Tanizaki:2015rda}, this structure is washed out if the contribution of only one thimble is taken into account.

\subsection{Real Time Dynamics}\label{sec:real}

%%%%%%% kedysh contour 

\begin{figure}[t] % no figure before 1st section
  \centering
  \includegraphics[width=8cm,angle=0,clip]{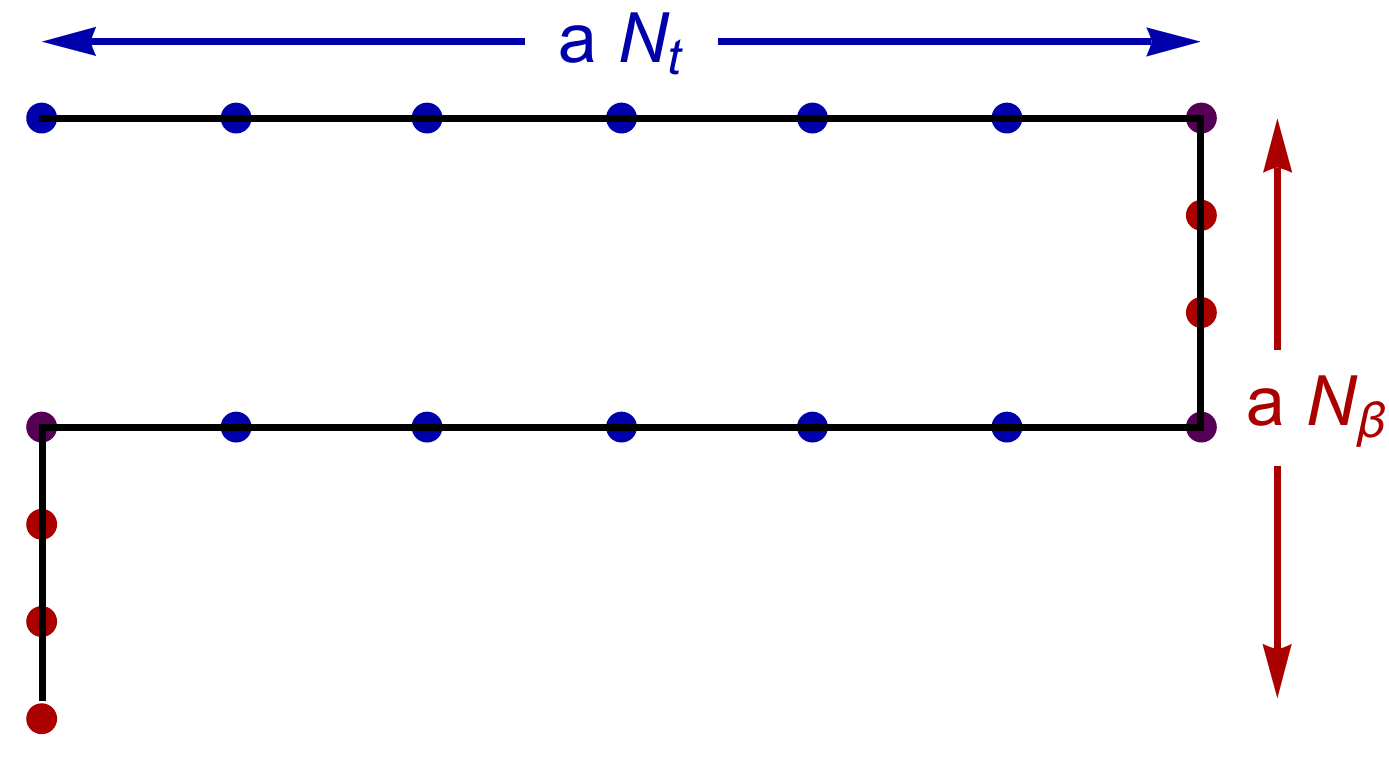}
  \caption{ 
  Contour in the complex $t$-plane used in the Schwinger-Keldysh formalism.
 }
  \label{fig:keldysh}% Give a unique label
\end{figure}
%%%%%%%%%%%%%%%%%%%
As is well known, a number of observables like particle masses and matrix elements, can be extracted  from the imaginary time correlators. Other observables like transport coefficients can be found only from real time correlators. While in thermal equilibrium the real time correlators can be, in principle, obtained from the knowledge of the imaginary time ones, the process requires an analytic continuation from the discrete, imaginary values of energy available in the Matsubara formalism to  continuous real energies. Although the analytic continuation  is possible (and unique), the analytic continuation is numerically unstable and the statistical uncertainty of a Monte Carlo calculations makes it nearly impossible in practice. An alternative approach would be a direct calculation in real time. A framework for calculations of thermal averages in real time exists, the so-called Schwinger-Keldysh formalism \cite{Schwinger:1960qe,Keldysh:1964ud}. It has been almost exclusively used in conjunction with weak coupling expansions (the exceptions we are aware of are the complex Langevin studies in \cite{Ilgenfritz:1986ca,Makri1988,Berges:2005yt,Berges:2006xc} ).

Consider a system described initially by a density matrix of the form $\rho=e^{-\beta \mathcal{H}}/Z$\footnote{There is no loss of generality assuming this form of $\rho$.} and whose evolution in time is governed by the hamiltonian $H$. 
We will be interested in observables of the form $\langle \mathcal{O}_1(t_1) \mathcal{O}_2(t_2) \rangle= {\rm tr}\left( 
\rho \mathcal{O}_1(t_1) \mathcal{O}_2(t_2)  \right)$. In the case $H=\mathcal{H}$ this is an equilibrium correlation function and it depends on  the difference $t_1-t_2$. Fully non-equilibrium situations can be dealt with  the same formalism but, for now, we will restrict ourselves to the equilibrium case.

Expectation value of observables are given by:
\bea
\langle \mathcal{O}(t)  \rangle
&=&
{\rm tr}\left( \rho \mathcal{O}(t)  \right)
=
{\rm tr}\left( \rho(0)U(t,0) \mathcal{O}U^\dagger(t,0) \right)\\
&=&
{\rm tr} \left(
U(-i\beta, -i\beta/2)U(-i\beta/2, T-i\beta/2)U(T-i\beta/2, T)U(T,t)\mathcal{O}U(t,0)
\right)
\eea where $U(t,t') = e^{-iH(t-t')}$ and $T$ is an arbitrary time $T>t$.  The case of 2-point functions or more is dealt with similarly. This expression describes the propagation from $0$ to $t$, then to $T$, down to $T-i\beta/2$, then  backwards to $-i\beta/2$ and down the along the imaginary direction to $-i\beta$. It has a path integral representation in terms of an action defined on a time contour in the complex $t$ plane depicted in fig.\ref{fig:keldysh} \footnote{There is a certain latitude in the choice of this contour. All that is required is that it starts at $t=0$, ends at $t=-i\beta$ a passes through $t$ on the real axis.}:
\beq\label{eq:sk}
\langle \mathcal{O}(t)  \rangle = \frac{
\int D\phi\ e^{-S_{SK}[\phi]} \mathcal{O}
}
{
\int D\phi\ e^{-S_{SK}[\phi]} 
}.
\eeq The contour action $S_c$ is purely imaginary along the part of the contour lying on the real axis. The discretization we used was:
\bea\label{eq:sk_discrete}
S_{SK}[\phi] 
= \sum_{t,n} a_t a  && \left[
 \frac{(\phi_{t+1,n}-\phi_{t,n})^2}{2 a_t^2} 
+ \frac12 \left( \frac{(\phi_{t+1,n+1}-\phi_{t+1,n})^2}{2 a^2} 
+  \frac{(\phi_{t,n+1}-\phi_{t,n})^2}{2 a^2} 
\right)   \right.\nonumber\\
&&+ \left.  \frac{m^2}{2} \frac{ \phi_{t,n}^2 + \phi_{t+1,n}^2}{2}  
+   \frac{\lambda}{4!} \frac{\phi_{t+1,n}^4 +  \phi_{t,n}^4 }{2}
\right] 
\eea where $t$ and $n$ indexes the lattice along the time and spatial directions,  $a$ is the spatial lattice spacing and $a_t$ is the time lattice spacing:
\bea
a_t &=& ia, \quad \text{for} \quad 0\leq t< N_t, \nonumber\\
a_t &=& a, \quad \text{for} \quad N_t\leq t< N_t+N_\beta/2, \nonumber\\
a_t &=& -ia,\quad \text{for} \quad  N_t+N_\beta/2\leq t< 2N_t+N_\beta/2,  \nonumber\\
a_t &=& a, \quad \text{for} \quad 2N_t+N_\beta/2\leq t< 2N_t+N_\beta,
\eea 
and $N_t, N_\beta$ are the number of lattice points on the real and imaginary axis, respectively.

The integral in \eq{sk} has ``perfect" sign problem: the average sign is not small but strictly zero, as can be seen by considering the variation of the contour action when the value of the field on a real value of $t$ is considered. The real part of the action is not changed by this variation and there is no damping of the Boltzmann factor as $\phi(t)$ is varied.

\begin{figure}[t]
\centering
\subfigure[]{
       \includegraphics[width=6cm,angle=0,clip]{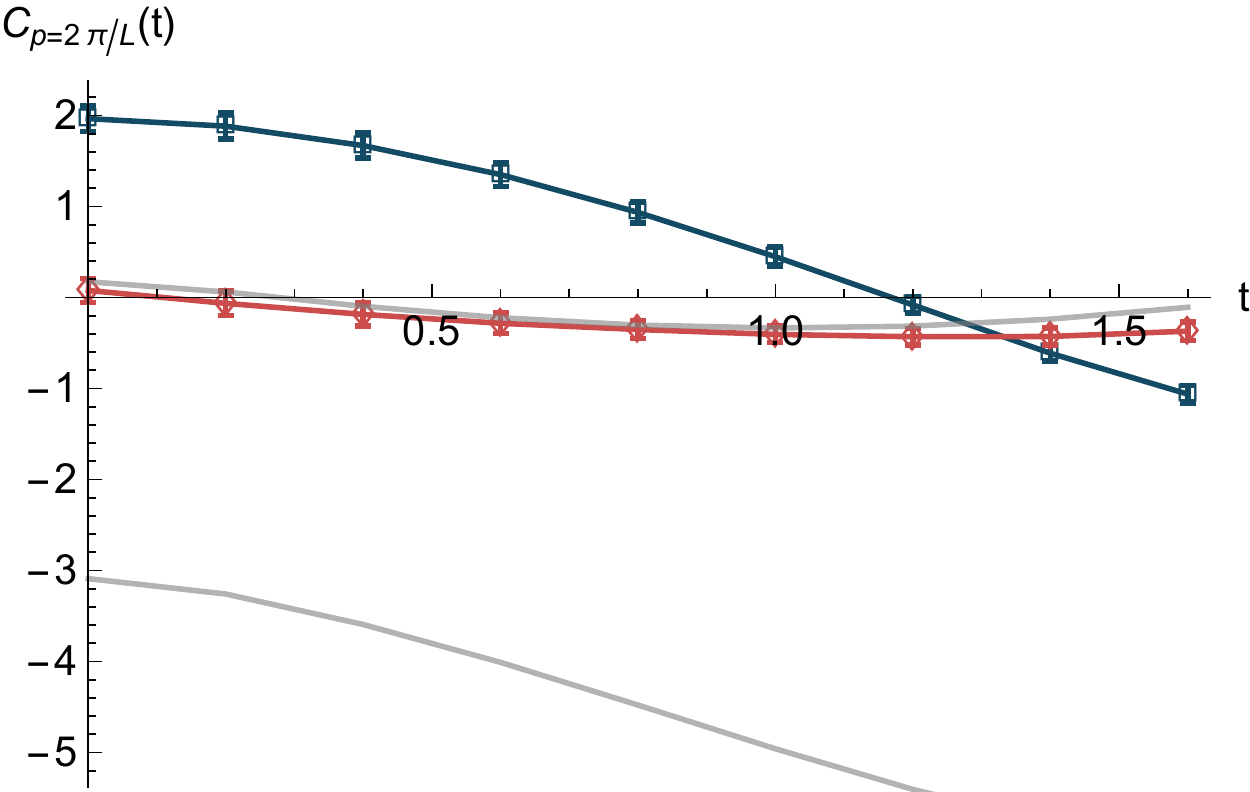}
     %  \label{fig:subfig1}
}
\subfigure[]{
    \includegraphics[width=6cm,angle=0,clip]{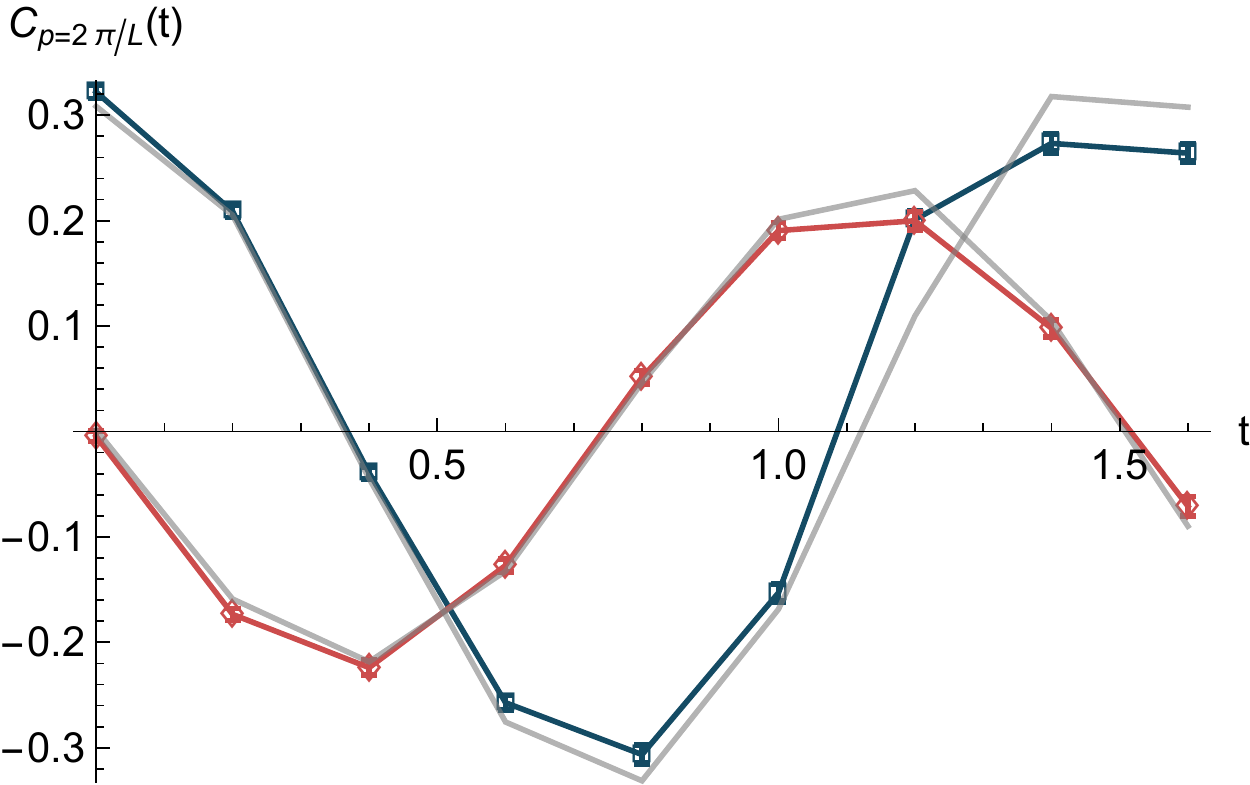}
    %\label{fig:subfig2}
}
\caption[ ]{Real (blue, squares) and imaginary (red, diamonds) parts of momentum projected correlators ($p=0$ on the left , $p=2\pi/L$ on the right)  for the parameters set: $N_x=N_t=8, N_\beta=2, m=1, a=0.2, \lambda=1$. The grey lines are the results from $\mathcal{O}(\lambda)$ perturbative calculations. The $p=0$ mode is clearly non-perturbative.}
\label{fig:real2}
\end{figure}

The application of the generalized thimble method to real time problems faces some extra difficulties. Consider, for instance, a $1+1$ dimensional $\phi^4$ theory \cite{Alexandru:2017lqr}. The first problem is that the tangent plane to the critical point $\phi_c =0$ is not real and, in some directions in field space, points towards the imaginary direction. Therefore, the tangent space is likely not in the same homology class as $\mathbb{R}^N$ and is not a suitable integration domain. The computationally cheap improvement of the sign problem obtained by simply integrating over the tangent space to the thimble is not available in this problem. We instead use the usual flow and perform the integration on the manifold $\mathcal{M}_T$ defined bey a substantial amount of flow. The leads to the problems listed in section \ref{sec:basic} being exacerbated. We use then the Grady method of section \ref{sec:pseudo} and validated by performing calculation at small $\lambda=0.1$ coupling and comparing it with the results from perturbation theory. For large values of the coupling the theory is no longer perturbative although the higher momentum modes still are.
We show  in fig.\ref{fig:real2}
some of the results for the parameter set $N_x=8, N_t=8, N_\beta=2, m=1, a=0.2$ and $\lambda=1$.

%%%%%%% real time results 

Our calculations were limited by the need to use very large flow times when $t$ is large. This leads to trapping of the Monte Carlo chain. Using a smaller flow time requires prohibitively large statistics. This is a problem that must be solved before the computation of transport coefficients becomes feasible.

\section{Future Prospects}\label{sec:future}

Considering the pace of development in the last couple of years, there is still plenty of room for algorithmic development in the thimble approach to the sign problem. Even the ideas discussed in this paper have not been explored fully; they should be applied to more models, implemented more efficiently, perhaps in more powerful computers (all calculations shown here were performed in run-of-the-mill laptops/desktops).  Certain features and trade-offs are only understood when larger, more realistic calculations are attempted. Even the models that were already studied were studied on a cursory way with the goal of developing techniques, not understanding the physics (no effort was put into approaching the continuum or large volume limits).  In addition, a few novel ideas were recently put forward and are still largely undeveloped. 

In order to bypass the computational cost of ``flowing" points from $\mathbb{R}^N$ to $\mathcal{M}_T$, it would be convenient to have a analytic parametrization of the manifold $\mathcal{M}_T$ and the associated jacobian. It is far from obvious how to find such a form. One possibility is to compute $\phi_T(\zeta)$ with a suitably large $T$ for a set of representative values of $\zeta$, interpolate them and then use the interpolating function as the definition of $\mathcal{M}_T$. This multi-dimensional interpolation is not an easy feat. In \cite{Alexandru:2017czx} this problem was attacked through the use of machine learning techniques. More precisely, a feed-forward network was setup that takes as an input the real values $\re \phi_T$ of the flowed field in $\mathcal{M}_T$ and  returns (an approximation to) the imaginary values of $\im\phi_T$. The map:
\beq\label{eq:learn}
\re\phi_i \rightarrow \re\phi_i + i\ \im\phi _i,
\eeq defines a manifold (the ``learnifold") $\mathcal{L}_T \approx \mathcal{M}_T$ that, after the network is ``trained", that is, its parameters are adjusted so to fit the know point obtained by flow,  becomes the new integration manifold. After the training process, new points on  $\mathcal{L}_T$ can be sampled at very low computational cost and large statistics can be easily accumulated. For periodic field variables, it is easy to see that the learnifold is a legitimate manifold of integration as it is continuously connected to the original domain. The parametrization in \eq{learn} has an additional advantage. The regions in $\re\phi$ mapped to the relevant regions in  $\mathcal{L}_T$ are not strechted out as in the parametrization of  $\mathcal{M}_T$ by the initial condition of the flow $\zeta$. Thus, the different modes of the distribution are not well separated and the Monte Carlo chain is much less likely to be trapped in a single mode.  Also, the absolute value of the jacobian does not fluctuate as much and it will be more easily reweigthed during measurements.

---------------------------------------

\section{Acknowledgements}
The research presented here was done in collaboration with Andrei Alexandru, Gokce Basar, Hank Lamm, Scott Lawrence, Greg Ridgway, Sohan Vartak and Neill Warrington. This research was supported by the U.S. Department of Energy under Contract No. DE-FG02-93ER-40762.

\clearpage
%\bibliography{lattice2017}
\bibliography{bedaque}

%%%%%%%%%%%%%%%%%%%%%%%%%%%%%%%%%%%%%%%%%%%%%%%%%%%%%%%%%%%%%%%%%%%%%%%%%%%%%
\end{document}